\documentclass[conference]{IEEEtran}
\IEEEoverridecommandlockouts
\usepackage{cite}
\ifCLASSINFOpdf
  \usepackage[pdftex]{graphicx}
\else
\fi
\usepackage{amsmath}

\usepackage{stfloats}

\usepackage{standalone}
\usepackage{tikz}
\usetikzlibrary{shapes,arrows}
\usepackage{bm,times}
\usepackage{mathtools}
\usepackage{booktabs}
\usepackage{subfigure}
\usepackage{enumitem}
\usepackage{layouts}
\usepackage{pgfplots}

\begin{document}
\title{Radar Pulse Deinterleaving with Transformer Based Deep Metric Learning\thanks{This work was supported by the Turing’s Defence and Security programme through a partnership with the UK government in accordance to the framework agreement between HMG \& The Alan Turing Institute.}}

\author{
\IEEEauthorblockN{
Edward Gunn\IEEEauthorrefmark{1}\IEEEauthorrefmark{3}, 
Adam Hosford\IEEEauthorrefmark{2},  
Daniel Mannion\IEEEauthorrefmark{1},
Jarrod Williams\IEEEauthorrefmark{1},
Varun Chhabra\IEEEauthorrefmark{1},
Victoria Nockles\IEEEauthorrefmark{1}
}
\IEEEauthorblockA{\IEEEauthorrefmark{1}DARe, The Alan Turing Institute \\ 
\IEEEauthorrefmark{2}DSTL \\
\IEEEauthorrefmark{3}Email: egunn@turing.ac.uk} 
}


% \author{\IEEEauthorblockN{Michael Shell\IEEEauthorrefmark{1},
% Homer Simpson\IEEEauthorrefmark{2},
% James Kirk\IEEEauthorrefmark{3}, 
% Montgomery Scott\IEEEauthorrefmark{3} and
% Eldon Tyrell\IEEEauthorrefmark{4}}
% \IEEEauthorblockA{\IEEEauthorrefmark{1}School of Electrical and Computer Engineering\\
% Georgia Institute of Technology,
% Atlanta, Georgia 30332--0250\\ Email: see http://www.michaelshell.org/contact.html}
% \IEEEauthorblockA{\IEEEauthorrefmark{2}Twentieth Century Fox, Springfield, USA\\
% Email: homer@thesimpsons.com}
% \IEEEauthorblockA{\IEEEauthorrefmark{3}Starfleet Academy, San Francisco, California 96678-2391\\
% Telephone: (800) 555--1212, Fax: (888) 555--1212}
% \IEEEauthorblockA{\IEEEauthorrefmark{4}Tyrell Inc., 123 Replicant Street, Los Angeles, California 90210--4321}}


\maketitle

\begin{abstract}
When receiving radar pulses it is common for a recorded pulse train to contain pulses from many different emitters.
The radar pulse deinterleaving problem is the task of separating out these pulses by the emitter from which they originated. Notably, the number of emitters in any particular recorded pulse train is considered unknown. In this paper, we define the problem and present metrics that can be used to measure model performance. We propose a metric learning approach to this problem using a transformer trained with the triplet loss on synthetic data. This model achieves strong results in comparison with other deep learning models with an adjusted mutual information score of 0.882.
\end{abstract}


% \fi
\IEEEpeerreviewmaketitle

\section{Introduction}
\label{sec:intro}

Radar pulse deinterleaving aims to separate out a train of radar pulses by the emitters from which they originated. 
We want to transform a single interleaved pulse train into many smaller deinterleaved pulse trains where each train contains all the pulses from a single emitter and only pulses from that emitter. 
A major challenge in this problem is that the number of emitters and hence the number of pulse trains we need to separate out is unknown.

Deinterleaving is a key part of the electronic warfare (EW) \cite{Wiley} pipeline, falling under the signal analysis stage. It aims to extract useful information from the detected signals, allowing for further analysis. 
It is much easier to analyse many smaller deinterleaved signals than a single large interleaved signal for tasks such as determining the radar's operating mode \cite{ModeRecognition}, specific emitter identification (SEI) \cite{SEI}, and operator intention identification \cite{IntentionIdentification}.

The main challenges faced in deinterleaving arise from congested and contested environments. In congested environments there is a large number of emitters often with significant overlap in parameter space, meaning it is difficult to differentiate between them. In contested environments we are faced with radars designed to be hard to detect. This can lead to dropped pulses due to low signal to noise ratio as well as high variance between pulses from the same emitter. Some examples of these are emitters that hop between frequencies and those with a variable pulse repetition interval (PRI), meaning there is varying time between the start of emitted pulses. This could result in a system reporting that there are more radars than are actually present.

Classically, deinterleaving was achieved through identifying the PRI for each emitter. Techniques for this include the cumulative difference histogram (CDIF) \cite{CDIF} and the sequential difference histogram (SDIF) \cite{SDIF}. These techniques work well when detecting pulses from emitters with a constant PRI. However, as the electromagnetic environment becomes more crowded we will require techniques that are capable of deinterleaving trains containing emitters which are more agile, often in an adversarial way.

\begin{figure}
    % \hspace*{-1cm}            
    % Define a few styles and constants
\tikzstyle{node}=[circle, minimum size=22pt,inner sep=0pt]%
\tikzstyle{comma node}=[scale=1.8]
\tikzstyle{bracket node}=[scale=3]
\tikzstyle{blue node}=[node, fill=blue!24]%
\tikzstyle{green node}=[node, fill=green!24]%
\tikzstyle{red node}=[node, fill=red!24]%
\tikzstyle{arr} = [draw, thick, double, double equal sign distance,-implies]%

\begin{tikzpicture}
    \node (ltb) [bracket node] at (-0.7,2) {$\big\{$};
    \node (rtb) [bracket node] at (8.7,2) {$\big\}$};
    \node (lbb) [bracket node] at (-1.15,0) {$\big\{$};
    \node (rbb) [bracket node] at (9.15,0) {$\big\}$};
    \node (bluelb) [bracket node] at (-0.9,0) {$\{$};
    \node (bluerb) [bracket node] at (3.1,0) {$\}$};
    \node (redlb) [bracket node] at (3.5,0) {$\{$};
    \node (redrb) [bracket node] at (6.5,0) {$\}$};
    \node (greenlb) [bracket node] at (6.9,0) {$\{$};
    \node (greenrb) [bracket node] at (8.9,0) {$\}$};
    % \node (arr) at (4,1) {$\boldsymbol{\Downarrow}$};
    \draw[arr] (4,1.2) -- (4,0.8) {};
    \node (comma1) [comma node] at (3.3,-0.4) {$,$};
    \node (comma1) [comma node] at (6.7,-0.4) {$,$};

    \foreach \i/ \pos/ \col in {
    {1/(0,2)/blue node}, 
    {2/(1,2)/red node},
    {3/(2,2)/blue node},
    {4/(3,2)/green node},
    {5/(4,2)/red node},
    {6/(5,2)/green node},
    {7/(6,2)/blue node},
    {8/(7,2)/blue node},
    {9/(8,2)/red node}}
        \node (x\i) [\col] at \pos {$x_\i$};

    \foreach \i/ \pos/ \col in {
    {1/(-0.4,0)/blue node}, 
    {2/(4,0)/red node},
    {3/(0.6,0)/blue node},
    {4/(7.4,0)/green node},
    {5/(5,0)/red node},
    {6/(8.4,0)/green node},
    {7/(1.6,0)/blue node},
    {8/(2.6,0)/blue node},
    {9/(6,0)/red node}}
        \node (y\i) [\col] at \pos {$x_\i$};
    
    % \foreach \i in {1,...,9}
    %     \draw[arr] (x\i.south) -- (y\i.north);
\end{tikzpicture}
    \caption{The radar pulse deinterleaving problem requires partitioning a set of radar pulses by the emitter from which the pulses originated. Here the colours represent the emitters and the features vectors $x_{i}$ denote PDWs. This can be approached using metric learning where we generate an embedding for each pulse in the context of the whole pulse train and cluster these embeddings using an off the shelf clusterer.}
    \label{fig:deinterleaving}
\end{figure}
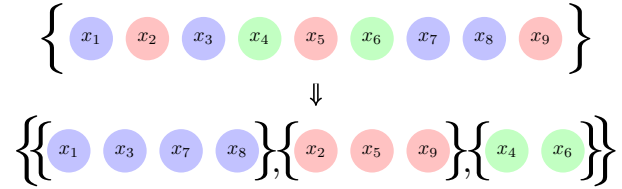

Recently the rise of deep learning has yielded new methods which can leverage large amounts of data to identify complex patterns and provide robustness to noise. The authors believe these approaches will be vital to successful deinterleaving in the presence of ever more agile digital radars. However, current applications of these methods have strict limitations. They can be largely divided into two categories. The first \cite{MultiparameterDeinterleaving, DotMatrix, SemanticSeg, ReconstructionRNN, DenoiseRNN, NMT, SepRefineNet, UNET} assumes that the number of emitters is fixed thus reducing the deinterleaving problem to a classification problem. This assumption is unlikely to hold in practice and the methods that make this assumption cannot easily adapt to different numbers of emitters. In this category a large number of works use variants of recurrent neural networks (RNNs) \cite{MultiparameterDeinterleaving, DotMatrix, SemanticSeg, ReconstructionRNN, DenoiseRNN, NMT} motivating our choice of baseline. Similar to classical deinterleavers, the second category of deep learning approaches to deinterleaving \cite{PRIModRec, DenoiseAE, HierarchicalDNN, MLDeinterleaving} attempt to identify pulse repetition intervals (PRIs) from times of arrival (ToAs) to deinterleave pulse trains. These methods often ignore additional pulse descriptor word (PDW) features extracted from raw IQ data, thereby restricting performance.

This paper seeks to address these limitations by proposing a pipeline that can leverage the full set of features while making no assumptions about the number of emitters.

Our key contributions are:

\begin{itemize}
\item We define the deinterleaving problem where the number of emitters is unknown and show how to measure performance with identified metrics (Section \ref{sec:problem-formulation}).
\item We combine metric learning with sequence-to-sequence models to propose a class of solutions to the problem as formulated. (Section \ref{subsec:metric-learning})
\item We analyse the types of model that would be most appropriate for solving this problem, concluding that the transformer is a strong candidate (Section \ref{subsec:model}).
\item We propose the use of the triplet loss with the batch all triplet mining strategy during training of the model (Section \ref{subsec:triplet}).
\item We train a sequence-to-sequence transformer with the triplet loss on a synthetic dataset and show that it has strong performance when compared to benchmark techniques (Section \ref{sec:experiement}).
\end{itemize}

\section{Problem formulation}
\label{sec:problem-formulation}

Given a \emph{pulse train} of $n$ radar pulses $X=\{ x_1, \dots, x_n\}$ from an unknown number of emitters $N$, we want to \emph{partition} the pulse train by the emitters from which the pulses originated. 
For example if the pulses $x_1,x_2,x_3,x_4,x_5$ came from emitters $1,1,2,2,1$ respectively the ground truth partition would be $\{\{x_1,x_2,x_5\},\{x_3,x_4\}\}$. The goal of deinterleaving is to create a model where for each pulse train in the test dataset the predicted partition $U=\{U_1, U_2,\dots \}$ is as similar as possible to the ground truth partition $V=\{V_1, V_2,\dots \}$, where $U_{i}$ and $V_{j}$ are the subsets of the partitions known as \emph{blocks}. To evaluate this we use standard \emph{extrinsic clustering} metrics and take their mean over every pulse train in the dataset. The metrics are:
\\[2ex]
\noindent{\bf Adjusted Mutual Information (AMI):}\\
 The \emph{mutual information} between two partitions of a set is
 \begin{equation}
     MI(U,V) = \sum_{i=1}^{|U|} \sum_{j=1}^{|V|} \frac{|U_{i} \cap V_{j}|}{n} \log \frac{n|U_{i} \cap V_{j}|}{|U_{i}||V_{j}|}
 \end{equation}
This does not adjust for chance so a random prediction will give a non-zero score. To account for this we use \emph{adjusted mutual information} \cite{AMI} which is given by
\begin{equation}
    AMI(U,V) = \frac{MI(U,V) - \mathbf{E}[MI(U,V)]}{\frac{1}{2}(H(U)+H(V))- \mathbf{E}[MI(U,V)]}
\end{equation}
where $n$ is the number of pulses in the train and $H(\cdot)$ is the entropy of a partition. 
AMI gives a value of 1 when the partitions are identical and 0 when the mutual information is equal to the expected mutual information.
\\[2ex]
\noindent{\bf Adjusted Rand Index (ARI):}\\
\emph{Rand index} \cite{ARI} considers pairs of points. A correct prediction is when a pair is predicted to be in the same block when they are truly in the same block (true positive) or if they are predicted to be in different blocks when they are truly in different blocks (true negative). We then define rand index as
\begin{equation}
    RI(U,V) = \frac{\text{\#correct predictions}}{\text{\#pairs}}
\end{equation}
This can be seen as the accuracy of a pairwise binary classification problem asking `are these two points from the same block'. Much like with AMI we adjust for chance to get the adjusted rand index with a value of 1 when the partitions are identical and 0 when the rand index is equal to the expected rand index.
\\[2ex]
\noindent{\bf V-measure:}\\
\emph{V-measure} \cite{Vmeasure} is defined as the harmonic mean of \emph{homogeneity} and \emph{completeness}. Homogeneity quantifies to what degree each cluster contains only members of a single class. Completeness quantifies to what degree all members of a given class are assigned to the same cluster. It is often useful to examine homogeneity and completeness individually to indicate whether a model has more of a tendency to split up points from the same ground truth cluster (low completeness) or to combine points from different ground truth clusters (low homogeneity). V-measure has a minimum value of 0 and a maximum value of 1 where higher is better. It is not adjusted for chance.
\\[2ex]

\section{Method}
\label{sec:method}

\subsection{A metric learning approach to deinterleaving}
\label{subsec:metric-learning}

Since the number of emitters is unknown, the number of blocks in the ground truth partition for any pulse train is also unknown. This means that familiar deep learning classification pipelines are not applicable. 

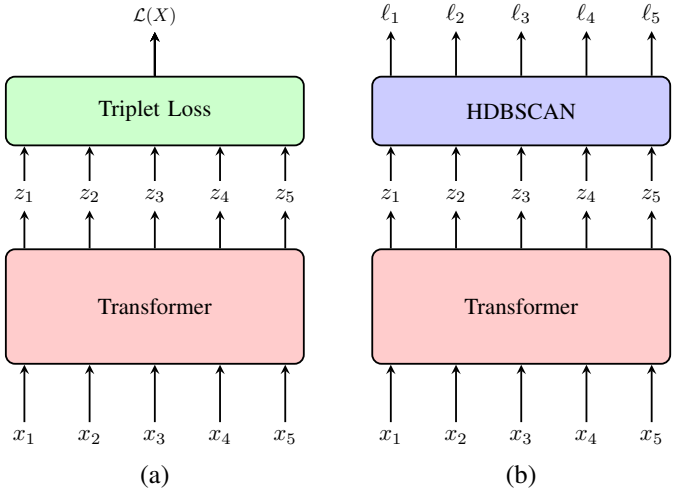
\begin{figure}[t]
	\centering
	\begin{minipage}[b]{.45\columnwidth}
		\centering
		% Define a few styles and constants
\def\numarrows{5}%

\tikzstyle{transformer} = [draw, text centered, fill=red!20, thick,
    minimum height=5em, minimum width=13em, rounded corners]%
\tikzstyle{loss} = [draw, text centered, fill=green!20, thick,
    minimum height=3em, minimum width=13em, rounded corners]%
\tikzstyle{arr} = [draw,thick,-stealth]%

\begin{tikzpicture}
    \node (transformer) [transformer] at (3,2) {Transformer};
    \node (loss) [loss] at (3,5) {Triplet Loss};
    \node[scale=0.8] (out) at (3,6.45) {$\mathcal{L}(X)$};

    \foreach \i in {1,...,\numarrows} {
      \node (x\i) at (\i,0) {$x_\i$};
      \node (z\i) at (\i,3.7) {$z_\i$};
      \draw [arr, stealth-] (transformer.south)+(\i-3,0) -- (x\i);
      \draw [arr] (transformer.north)+(\i-3,0) -- (z\i);
      \draw [arr, stealth-] (loss.south)+(\i-3,0) -- (z\i);
      \draw [arr] (loss.north) -- (out);
    }
  
\end{tikzpicture}
            (a)
	\end{minipage}%
        \hfill
	\begin{minipage}[b]{.45\columnwidth}
		\centering
		% Define a few styles and constants
\def\numarrows{5}%

\tikzstyle{transformer} = [draw, text centered, fill=red!20, thick,
    minimum height=5em, minimum width=13em, rounded corners]%
\tikzstyle{loss} = [draw, text centered, fill=blue!20, thick,
    minimum height=3em, minimum width=13em, rounded corners]%
\tikzstyle{arr} = [draw,thick,-stealth]%

\begin{tikzpicture}
    \node (transformer) [transformer] at (3,2) {Transformer};
    \node (loss) [loss] at (3,5) {HDBSCAN};

    \foreach \i in {1,...,\numarrows} {
      \node (x\i) at (\i,0) {$x_\i$};
      \node (z\i) at (\i,3.7) {$z_\i$};
      \node (l\i) at (\i,6.5) {$\ell_\i$};
      \draw [arr, stealth-] (transformer.south)+(\i-3,0) -- (x\i);
      \draw [arr] (transformer.north)+(\i-3,0) -- (z\i);
      \draw [arr, stealth-] (loss.south)+(\i-3,0) -- (z\i);
      \draw [arr] (loss.north)+(\i-3,0) -- (l\i);
    }
  
\end{tikzpicture}
            (b)
	\end{minipage}
        \caption{The architecture for the training (a) and inference (b) pipelines share the common component of a sequence-to-sequence transformer used as a pulse embedding model. We use vanilla dot product attention and no positional encodings. (a) During training we use the triplet loss to push apart embeddings of pulses from different emitters and bring together those from the same emitter. (b) During inference we use a clustering algorithm (HDBSCAN) to generate an integer label $l_i$ for each embedding from which we can trivially determine the predicted partition for the set of pulses.}
        \label{fig:pipelines}
\end{figure}

Instead we take a \emph{metric learning} \cite{MetricLearning} approach. We use a \emph{sequence-to-sequence} model to generate an \emph{embedding} $z_i$ for each pulse $x_i$ in a pulse train $X$. We train this model so that embeddings for pulses from the same emitter are close together and those from different emitters are far apart (as measured by their Euclidean distance).
During inference we can cluster these embeddings using a non-parametric clustering algorithm such as HDBSCAN \cite{HDBSCAN} to form a predicted partition for the original pulse train. A visual representation of the training and inference pipelines can be seen in Figure \ref{fig:pipelines}.

% \begin{figure}[ht]
%     \includestandalone[width=\columnwidth]{Figures/inference_pipeline}
%     \label{fig:inference-pipeline}
%     \caption{The inference pipeline blah blah}
% \end{figure}

For the deinterleaving problem individual pulses have little meaning in isolation. It is therefore important to use a sequence-to-sequence model to ensure that their embeddings are generated in the \emph{context} of the rest of the pulse train. In other words, the model must consider the \emph{relationships} between the pulses.

% It is important to note that this problem is subtly different from traditional clustering problems due to the clustering only being evaluated internally to each pulse train rather than across the whole dataset. This means the task we are giving the model is: \\
% \noindent{\textit{Given a set of samples, output an embedding for each sample that, when clustered, partitions the set},} \\ \noindent{rather than:} \\\noindent{\textit{Given a single sample, output an embedding that is a good representation for that sample in the space that has been learned},} \\ \noindent{as would be seen in most clustering problems.}

% The key difference between these two setups is context. For traditional clustering problems the context is baked into the model during training allowing us to use models that only take a single sample, in this case a pulse, as input.
% However, this implies relationships between samples are not considered.
% Since our problem is fundamentally relational this approach would be pathologically constrained.
% Instead, along with each input sample we must also include some context as a function of the other samples in the set.
% Without this context a vast amount of data would be necessary for training and the ability for a model to generalise would likely be severely limited.

\subsection{Transformer embedding model}
\label{subsec:model}

The transformer \cite{Transformer} is a popular architecture that has demonstrated state of the art results across natural language processing and computer vision. One of the key features that makes it powerful is the way it explicitly models the relationship between every pair of tokens through its attention mechanism. Every token can influence every other token and the strength and form of this influence is learned during training. 
Given the importance of relationships between pulses in the deinterleaving problem, as discussed in the previous section, the transformer's explicit modelling of these relationships makes it a compelling choice for our model.

Specifically, the model we consider is a vanilla sequence-to-sequence transformer with dot product self-attention. The order of detection of the pulses may be inferred by their ToA meaning positional encodings are not necessary and thus were not used. Preliminary experiments showed they had a small negative effect on performance, likely due to redundant information.

\subsection{Clustering embeddings with the triplet loss}
\label{subsec:triplet}

As stated in Section \ref{subsec:metric-learning} we want to train the model so that, within a pulse train, embeddings for pulses from
the same emitter are close together and those from different
emitters are far apart.
The triplet loss \cite{TripletLoss} is a common loss used for deep metric learning. The loss considers triplets of embeddings, an anchor $z$, a positive embedding $z^{+}$ which is from the same emitter as the anchor, and a negative embedding $z^{-}$ which is from a different emitter to the anchor. It compares the distance between the anchor and the positive to the distance between the anchor and the negative. It aims to satisfy the triplet inequality stated in equation \ref{eq:triplet-inequality}.
\begin{equation}
\label{eq:triplet-inequality}
d(z,z^{+}) + \alpha < d(z,z^{-})
\end{equation}
This says that we want the distance between the anchor and the positive to be less than the distance between the anchor and the negative by some margin $\alpha$ so that the embedding of pulses from the same emitter are closer together than those from different emitters. When this inequality is satisfied we do not want to update the model, so the loss must be 0. When it is not satisfied we want to penalise the model by the degree to which it is violated. We therefore use the hinge loss in equation \ref{eq:triplet-loss}.

% The triplet loss \cite{TripletLoss} is a technique used in metric learning which encourages precisely this across a whole dataset. 
% Using labeled data we generate an embedding for three samples. 
% These are the anchor $z$, the positive $z^{+}$, which comes from the same cluster as the anchor, and the negative $z^{-}$, which comes from a different cluster to the anchor.
% The loss is then calculated as 
\begin{equation}
    \label{eq:triplet-loss}
    \mathcal{L}(z,z^{+},z^{-}) = \max \{ d(z,z^{+})-d(z,z^{-})+\alpha,0\}
\end{equation}

% \noindent{where $d$ is a distance metric and $\alpha$ is a hyperparameter referred to as the \emph{margin}. }
% The margin is the distance beyond which there is no benefit in separating points from different clusters further; this is enforced by the $\max$ term in the loss. 

For each pulse train there is on the order of $\mathcal{O}(n^3)$ possible triplets where $n$ is the number of pulses in the pulse train. Often most of these triplets are relatively uninformative to the model. For example, easy triplets which satisfy equation \ref{eq:triplet-inequality} have a loss of $0$. Therefore, training using all possible triplets is very inefficient. 

To increase the efficiency of training with the triplet loss we need a method to choose triplets that will be informative for training. This is called a triplet mining strategy. Our experiments found that the triplet mining strategy with the best performance was the batch all \cite{FaceNet} strategy where we take the mean of the loss across all non-easy triplets.

The batch all triplet loss is defined as follows. Given the embeddings $Z = \{z_1, \dots , z_n \}$, let the set of all non-easy triplets be 
\begin{equation*}
    \mathcal{T}^+(Z)= \left\{ (z_i,z_j,z_k) \in Z^3 \middle\vert
    \begin{array}{cc}
         i \sim j \not\sim k, \\
         i \neq j \neq k, \\
         d(z_i,z_j) + \alpha \geq d(z_i,z_k)
    \end{array}
    \right\}
\end{equation*}
where $\sim$ is the equivalence relation where $i \sim j$ indicates that the pulses at index $i$ and $j$ come from the same emitter. The batch all triplet loss is

\begin{equation*}
  \mathcal{L}(Z)= \: \frac{1}{|\mathcal{T}^+(Z)|}\smashoperator{\sum_{\overset{\vphantom{c}}{(z_i,z_j,z_k) \in \mathcal{T}^+(Z)}}} \: \max \{ d(z_i,z_j)-d(z_i,z_k)+\alpha,0\}
\end{equation*}

In this work we use the Euclidean distance metric.

\section{Experimental work}
\label{sec:experiement}

\subsection{Datasets}
\label{subsec:datasets}

Using synthetic data is overwhelmingly the norm in deinterleaving. The authors are only aware of one paper \cite{MLDeinterleaving} that uses real data. This is for two main reasons. The first is that real data is very difficult to ground truth. Usually labels are assigned to pulses by an existing deinterleaver. Training on this data would result in the performance of our models converging to that of the existing deinterleavers. Since the goal of this work is to develop models that outperform existing deinterleavers this approach would be pathologically constrained. The second reason is that as far as the authors are aware there are no real publicly available datasets suitable for deinterleaving.

We generated synthetic training, validation, and test sets using a simulator. These contain $100,000$ pulse trains in the train set, $10,000$ in the validation set, and another $10,000$ pulse trains in the test set. Each pulse train consists of $1000$ pulses, and contains between $2$ and $20$ emitters. Although processing pulse trains with only one emitter in them is of interest, when this is the case no valid triplets can be formed. Therefore, training on them results in zero loss and the model is not updated.

\begin{figure*}
    \centering
    \input{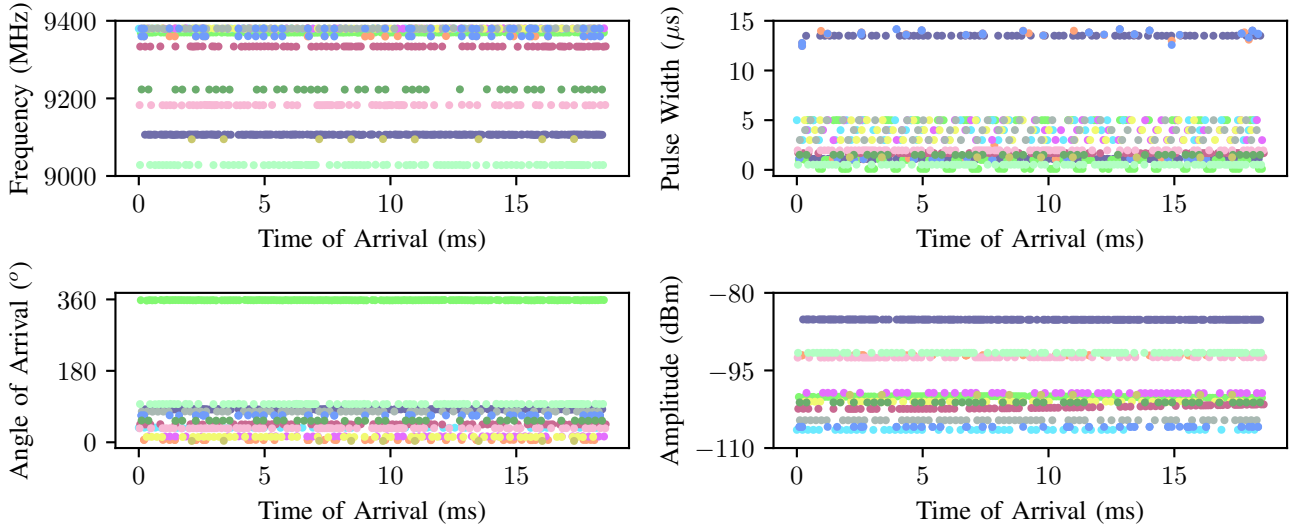}
    \caption{A plot of the different PDW features against time of arrival for an example pulse train in the test dataset. The pulse train contains 13 emitters with significant overlap in parameters. The colour of each data point indicates which of the emitters the pulse originated from.}
    \label{fig:pdws}
\end{figure*}

For feature vectors to represent pulses we use pulse descriptor words (PDWs) with 5 dimensions: time of arrival (ToA), centre frequency, pulse width (PW), angle of arrival (AoA), and amplitude. Figure \ref{fig:pdws} shows an example of the PDWs from one of the pulse trains in our test set.

% We generated a set of synthetic data using a simulator. This dataset contains 356 pulse trains of length varying between 294,190 and 2,087,694 pulses, each containing up to 11 emitters. These trains were then distributed in the ratio 8:1:1 among the train, validation, and test sets.
% Each train is split into subtrains containing 1000 pulses. Any remaining pulses are discarded. For trains with only 1 emitter it is not possible to form any valid triplets resulting in no loss contribution from these trains. We therefore also discard any subtrains that only contain pulses from a single emitter.
% Once this preprocessing is completed we are left with a training dataset containing 288,399 trains of 1000 PDWs.

% Additionally we used a small real dataset collected in a maritime environment. [CONFIRM PASSGE BEFORE ADDING HERE]

% ADD VISUALISATION OF DATA HERE

\subsection{Model description}
\label{subsec:parameters}

We use 8 transformer layers with 8 attention heads, a feed-forward hidden size of 2048, and a residual size of 256. The model is trained with Adam \cite{Adam} with a learning rate of 0.0001, and a batch size of 8 for 8 epochs. The triplet loss margin $\alpha$ is 1.9, and dropout is 0.05. Embeddings $z_i$ are projected to 8 dimensions. Hyperparameters were selected via a parameter sweep on the validation set.

\subsection{Data Normalization}
\label{subsec:normalisation}

We normalise the data internally to each pulse train only. To do this we treat each element of the PDWs separately. The ToAs are linearly re-scaled so that the lowest ToA is 0 and the highest is 1. The centre frequency, pulse width, and amplitude are all statistically normalised by taking away the mean and dividing by the standard deviation in the pulse train. Finally AoA is divided by $360^{\circ}$ to place it in the range $[0,1]$.

\subsection{Baselines}
\label{subsec:baseline}

As far as we are aware there are no readily available deep learning based benchmarks for this problem as formulated.
In addition to this, the datasets on which other results were tested are not publicly available making comparison between solutions challenging.

Instead, to gain a performance baseline we trained a sequence-to-sequence gated recurrent unit (GRU) \cite{GRU} model using the same triplet loss discussed in Section \ref{subsec:triplet}. Similar architectures to this are widely used in deinterleaving related literature as discussed in Section \ref{sec:intro}. The key point to note about the GRU is that the relationships between pulses are expressed less explicitly than in the transformer. An internal state vector is propagated through the network meaning tokens only attend to those before it in the sequence.
We used 8 GRU layers with a hidden size of 512 resulting in the number of parameters being roughly the same as that of the transformer.

As a sanity check we also include results obtained using an identity model where the PDWs themselves are clustered using HDBSCAN. We found the minimum cluster size parameter of HDBSCAN had a large effect on performance. For our experiments we set it to 20 for all models.

\begin{figure}
    \centering
    \input{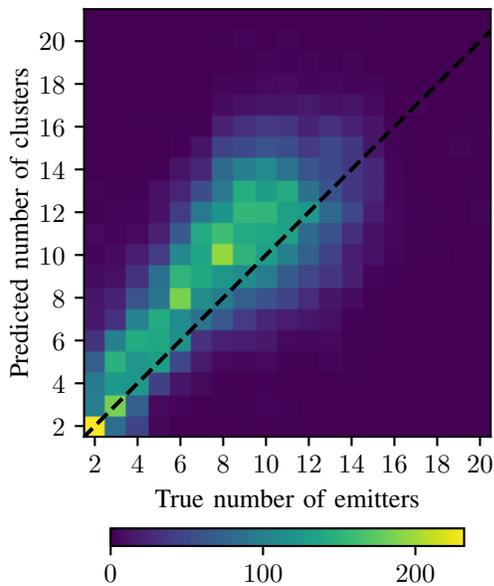}
    \caption{A confusion matrix of the true number of emitters in the synthetic test set against the number of clusters in the partition predicted by our transformer. The black line represents an ideal scenario where the predicted number of clusters is the same as the true number of clusters. It can be seen that the transformer generally predicts there to be more clusters than there are emitters.}
    \label{fig:confusion-matrix}
\end{figure}

\begin{table}[b]
% \renewcommand{\arraystretch}{1}
% if using array.sty, it might be a good idea to tweak the value of
% \extrarowheight as needed to properly center the text within the cells
\caption{Results on the test set}
\label{tab:results}
\centering
% Some packages, such as MDW tools, offer better commands for making tables
% than the plain LaTeX2e tabular which is used here.
\begin{tabular}{c|ccc}
Deinterleaver & Transformer & GRU & Identity\\ 
        \hline
        AMI & \textbf{0.882} & 0.850 & 0.761 \\
        ARI & \textbf{0.817} & 0.811 & 0.610 \\
        V-measure & \textbf{0.884} & 0.852 & 0.765 \\
        Homogeneity & 0.938 & 0.856 & 0.899 \\
        Completeness & 0.853 & 0.872 & 0.697 \\
\end{tabular}
\end{table}

\subsection{Results and discussion}
\label{subsec:results}

As shown in Table \ref{tab:results}, the transformer outperforms our chosen baseline models in all performance metrics from Section \ref{sec:problem-formulation}. 

Figure \ref{fig:confusion-matrix} displays the results of the transformer when predicting the number of clusters in each pulse train. These values are extracted directly from the deinterleaved pulse trains by counting the number of clusters present. This, along with the homogeneity and completeness scores, indicates that the transformer and identity models have a general trend to overestimate the number of clusters, with a RMS error of $3.07$, and $3.48$ respectively, whereas the GRU tends to underestimate the number of clusters, with an RMS error of $2.57$. The transformer achieves a completeness score of $0.938$ meaning clusters overwhelmingly contain only pulses from a single emitter. From our perspective, correctly forming clusters that contain pulses from only one emitter, even if we incorrectly split that emitter's pulses into separate clusters, is better than mixing pulses from different emitters together. This is because it suggests the deinterleaver is very effective at distinguishing between different emitters. Furthermore, we believe this incorrect splitting of pulses from a single emitter into multiple clusters could be corrected during a post processing step which combines related clusters - thus increasing completeness and V-measure.

\begin{figure}
    \input{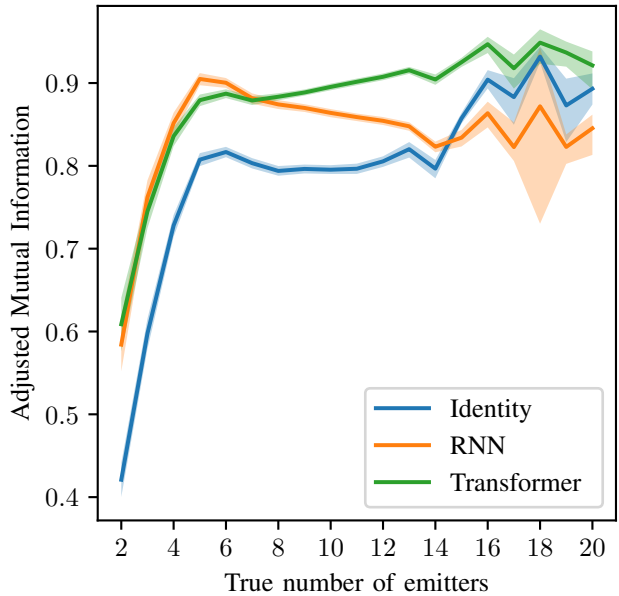}
    \caption{The true number of emitters against the mean AMI for that number of clusters. The shaded area represents the bootstrapped \cite{Bootstrap} $0.1-0.9$ confidence interval. It is notable that the width of the confidence interval is much larger for high numbers of clusters where we have few data points (The distribution of data points can be seen in Figure \ref{fig:confusion-matrix}).}
    \label{fig:cluster-num-metrics}
\end{figure}

\begin{figure}
    \hspace{0.5cm}\input{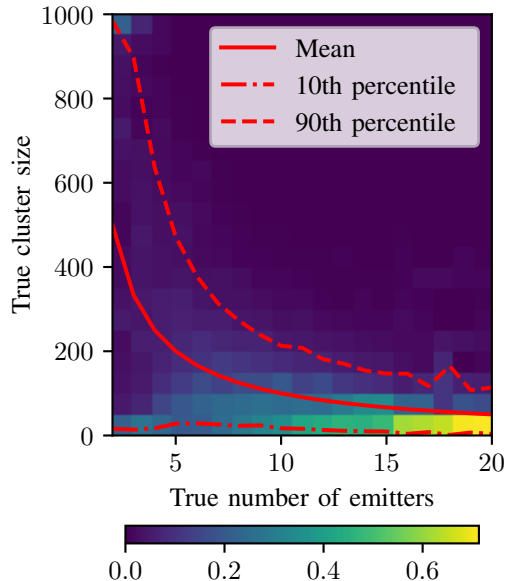}
    \caption{The true number of emitters against the true cluster size (the number of pulses from each emitter in a pulse train). The lines represent the mean, 10th percentile and 90th percentile. They show that the size and variability in size of clusters decreases as the number of emitters increases. The heat map colours represent the proportion of clusters in pulse trains with a particular number of emitters that are within a given size bin. They show that for low numbers of emitters there is significant imbalance between cluster sizes, as the number of emitters increases the cluster sizes become more balanced.}
    \label{fig:cluster-size-distribution}
\end{figure}

Figure \ref{fig:cluster-num-metrics} shows the mean adjusted mutual information for models across an increasing number of emitters. The performance is strong for 5 or more emitters. However, when there are few ground truth clusters performance is poor, regardless of the model used. This suggests the issue is not related to the training scheme, as the untrained identity model exhibits a similar trend.
% The performance increases rapidly from 2 to 5 ground truth clusters, then largely plateaus. This indicates the effect of the interplay between learning on adjacent numbers of emitters is not a major factor. 
The problem must therefore be due to the dataset or the clustering algorithm itself.
All the models tend to over-split clusters when the number of emitters is low. They maintain high homogeneity, but completeness drops rapidly for fewer than 5 emitters. This could be due to three reasons, each identifiable in Figure \ref{fig:cluster-size-distribution}:

\begin{enumerate}
    \item The ground truth clusters are generally much larger for low numbers of emitters.
    \item The variability between ground truth cluster sizes is much higher for low numbers of emitters.
    \item Imbalance between cluster sizes is more prominent for small numbers of emitters, with one big cluster and many smaller ones.
\end{enumerate}

All these effects are likely amplified by the fact that pulse trains have been limited to length 1000.
By addressing these potential issues in the dataset or clustering approach, the performance for low numbers of emitters could likely be improved.

% \begin{table}[ht]
%     \centering
%     \begin{tabular}{c|ccc}
%         Deinterleaver & Transformer & GRU & Identity\\ 
%         \hline
%         AMI & \textbf{0.882} & 0.850 & 0.761 \\
%         ARI & \textbf{0.817} & 0.811 & 0.610 \\
%         V-measure & \textbf{0.884} & 0.852 & 0.765 \\
%         Homogeneity & 0.938 & 0.856 & 0.899 \\
%         Completeness & 0.853 & 0.872 & 0.697 \\
%         \end{tabular}
%     \caption{The results on the test set for the transformer against the baseline of a gated recurrent unit and a identity map where we cluster the PDWs without any transformation. The transformer demonstrates the strongest performance over all metrics from Section \ref{sec:problem-formulation}.}
%     \label{tab:results}
% \end{table}

\section{Future work}
\label{sec:related-work}

Future work will explore addressing the issues concerning performance at low numbers of emitters as discussed in Section \ref{subsec:results}. Equally, scaling this model to longer and variable length sequences would also prove beneficial as the dynamics of the previously discussed clipping issues could be reduced. In practice, real pulse trains are longer and variable in length so this scenario would be more realistic.
% This will require exploring models and losses with lower computational complexity than the transformer and triplet loss. 
Finally, we will explore applying this model to real data collects. The main challenge involved in this is obtaining reliable ground truth labels as discussed in section \ref{subsec:datasets}.
% Finally PDWs are only one possible representation of radar pulses and future work could explore enriching them through learned features on the raw IQ data.

\section{Conclusion}
\label{sec:conclusion}

In this paper we defined the deinterleaving problem and proposed a metric learning approach as a solution using a transformer trained with the triplet loss on a synthetic dataset. We showed that it achieves strong results in comparison with an identity baseline indicating that this method is effective for solving the deinterleaving problem. Additionally we showed that, in this pipeline, the transformer architecture outperforms a GRU architecture commonly used in the literature. Future work on transformers with this pipeline can be exploited to further improve deinterleaving performance.

% \section*{Acknowledgment}




\bibliographystyle{IEEEtran}
\bibliography{refs.bib}

\begin{thebibliography}{10}
\providecommand{\url}[1]{#1}
\csname url@samestyle\endcsname
\providecommand{\newblock}{\relax}
\providecommand{\bibinfo}[2]{#2}
\providecommand{\BIBentrySTDinterwordspacing}{\spaceskip=0pt\relax}
\providecommand{\BIBentryALTinterwordstretchfactor}{4}
\providecommand{\BIBentryALTinterwordspacing}{\spaceskip=\fontdimen2\font plus
\BIBentryALTinterwordstretchfactor\fontdimen3\font minus \fontdimen4\font\relax}
\providecommand{\BIBforeignlanguage}[2]{{%
\expandafter\ifx\csname l@#1\endcsname\relax
\typeout{** WARNING: IEEEtran.bst: No hyphenation pattern has been}%
\typeout{** loaded for the language `#1'. Using the pattern for}%
\typeout{** the default language instead.}%
\else
\language=\csname l@#1\endcsname
\fi
#2}}
\providecommand{\BIBdecl}{\relax}
\BIBdecl

\bibitem{Wiley}
R.~Wiley, \emph{ELINT: The interception and analysis of radar signals}.\hskip 1em plus 0.5em minus 0.4em\relax Artech, 2006.

\bibitem{ModeRecognition}
\BIBentryALTinterwordspacing
J.~Ou, Y.~Chen, F.~Zhao, J.~Liu, and S.~Xiao, ``Novel approach for the recognition and prediction of multi-function radar behaviours based on predictive state representations,'' \emph{Sensors}, vol.~17, no.~3, 2017. [Online]. Available: \url{https://www.mdpi.com/1424-8220/17/3/632}
\BIBentrySTDinterwordspacing

\bibitem{SEI}
A.~Kawalec and R.~Owczarek, ``Radar emitter recognition using intrapulse data,'' in \emph{15th International Conference on Microwaves, Radar and Wireless Communications (IEEE Cat. No. 04EX824)}, vol.~2.\hskip 1em plus 0.5em minus 0.4em\relax IEEE, 2004, pp. 435--438.

\bibitem{IntentionIdentification}
J.~Liang, B.~I. Ahmad, M.~Jahangir, and S.~Godsill, ``Detection of malicious intent in non-cooperative drone surveillance,'' in \emph{2021 Sensor Signal Processing for Defence Conference (SSPD)}, 2021, pp. 1--5.

\bibitem{CDIF}
H.~Mardia, ``New techniques for the deinterleaving of repetitive sequences,'' in \emph{IEE Proceedings F (Radar and Signal Processing)}, vol. 136, no.~4.\hskip 1em plus 0.5em minus 0.4em\relax IET, 1989, pp. 149--154.

\bibitem{SDIF}
Z.~Ge, X.~Sun, W.~Ren, W.~Chen, and G.~Xu, ``Improved algorithm of radar pulse repetition interval deinterleaving based on pulse correlation,'' \emph{IEEE Access}, vol.~7, pp. 30\,126--30\,134, 2019.

\bibitem{MultiparameterDeinterleaving}
W.~Chao, L.~Weisong, L.~Xueqiong, W.~Xiang, and H.~Zhitao, ``A new radar signal multiparameter-based deinterleaving method,'' 2022.

\bibitem{DotMatrix}
\BIBentryALTinterwordspacing
A.~Al-Malahi, A.~Farhan, H.~Feng, O.~Almaqtari, and B.~Tang, ``An intelligent radar signal classification and deinterleaving method with unified residual recurrent neural network,'' \emph{IET Radar, Sonar \& Navigation}, vol.~17, no.~8, pp. 1259--1276, 2023. [Online]. Available: \url{https://ietresearch.onlinelibrary.wiley.com/doi/abs/10.1049/rsn2.12417}
\BIBentrySTDinterwordspacing

\bibitem{SemanticSeg}
\BIBentryALTinterwordspacing
W.~Chao, S.~Liting, L.~Zhangmeng, and H.~Zhitao, ``A radar signal deinterleaving method based on semantic segmentation with neural network,'' \emph{IEEE Transactions on Signal Processing}, vol.~70, p. 5806–5821, 2022. [Online]. Available: \url{http://dx.doi.org/10.1109/TSP.2022.3229630}
\BIBentrySTDinterwordspacing

\bibitem{ReconstructionRNN}
\BIBentryALTinterwordspacing
H.~Zheng, K.~Xie, Y.~Zhu, J.~Lin, and L.~Wang, ``An reconstruction bidirectional recurrent neural network ‐based deinterleaving method for known radar signals in open‐set scenarios,'' \emph{IET Radar, Sonar \& Navigation}, 2024. [Online]. Available: \url{https://api.semanticscholar.org/CorpusID:267469824}
\BIBentrySTDinterwordspacing

\bibitem{DenoiseRNN}
Z.-M. Liu and P.~S. Yu, ``Classification, denoising, and deinterleaving of pulse streams with recurrent neural networks,'' \emph{IEEE Transactions on Aerospace and Electronic Systems}, vol.~55, no.~4, pp. 1624--1639, 2019.

\bibitem{NMT}
M.~Zhu, S.~Wang, and Y.~Li, ``Model-based representation and deinterleaving of mixed radar pulse sequences with neural machine translation network,'' \emph{IEEE Transactions on Aerospace and Electronic Systems}, vol.~58, no.~3, pp. 1733--1752, 2022.

\bibitem{SepRefineNet}
\BIBentryALTinterwordspacing
Y.~Mao, W.~Ren, X.~Li, Z.~Yang, and W.~Cao, ``Sep-refinenet: A deinterleaving method for radar signals based on semantic segmentation,'' \emph{Applied Sciences}, vol.~13, no.~4, 2023. [Online]. Available: \url{https://www.mdpi.com/2076-3417/13/4/2726}
\BIBentrySTDinterwordspacing

\bibitem{UNET}
Z.~Kang, Y.~Zhong, Y.~Wu, and Y.~Cai, ``Signal deinterleaving based on u-net networks,'' in \emph{2023 8th International Conference on Computer and Communication Systems (ICCCS)}, 2023, pp. 62--67.

\bibitem{PRIModRec}
J.-W. Han and C.~H. Park, ``A unified method for deinterleaving and pri modulation recognition of radar pulses based on deep neural networks,'' \emph{IEEE Access}, vol.~9, pp. 89\,360--89\,375, 2021.

\bibitem{DenoiseAE}
X.~Li, Z.~Liu, and Z.~Huang, ``Deinterleaving of pulse streams with denoising autoencoders,'' \emph{IEEE Transactions on Aerospace and Electronic Systems}, vol.~56, no.~6, pp. 4767--4778, 2020.

\bibitem{HierarchicalDNN}
Z.-M. Liu, ``Pulse deinterleaving for multifunction radars with hierarchical deep neural networks,'' \emph{IEEE Transactions on Aerospace and Electronic Systems}, vol.~57, no.~6, pp. 3585--3599, 2021.

\bibitem{MLDeinterleaving}
\BIBentryALTinterwordspacing
H.~Wang, M.~Zhu, R.~Fan, and Y.~Li, ``Parametric model-based deinterleaving of radar signals with non-ideal observations via maximum likelihood solution,'' \emph{IET Radar, Sonar \& Navigation}, vol.~16, no.~8, pp. 1253--1268, 2022. [Online]. Available: \url{https://ietresearch.onlinelibrary.wiley.com/doi/abs/10.1049/rsn2.12258}
\BIBentrySTDinterwordspacing

\bibitem{AMI}
\BIBentryALTinterwordspacing
N.~X. Vinh, J.~Epps, and J.~Bailey, ``Information theoretic measures for clusterings comparison: Variants, properties, normalization and correction for chance,'' \emph{Journal of Machine Learning Research}, vol.~11, no.~95, pp. 2837--2854, 2010. [Online]. Available: \url{http://jmlr.org/papers/v11/vinh10a.html}
\BIBentrySTDinterwordspacing

\bibitem{ARI}
\BIBentryALTinterwordspacing
W.~M. Rand, ``Objective criteria for the evaluation of clustering methods,'' \emph{Journal of the American Statistical Association}, vol.~66, no. 336, pp. 846--850, 1971. [Online]. Available: \url{https://www.tandfonline.com/doi/abs/10.1080/01621459.1971.10482356}
\BIBentrySTDinterwordspacing

\bibitem{Vmeasure}
\BIBentryALTinterwordspacing
A.~Rosenberg and J.~Hirschberg, ``V-measure: A conditional entropy-based external cluster evaluation measure,'' in \emph{Conference on Empirical Methods in Natural Language Processing}, 2007. [Online]. Available: \url{https://api.semanticscholar.org/CorpusID:14153811}
\BIBentrySTDinterwordspacing

\bibitem{MetricLearning}
M.~Kaya and H.~{\c{S}}. Bilge, ``Deep metric learning: A survey,'' \emph{Symmetry}, vol.~11, no.~9, p. 1066, 2019.

\bibitem{HDBSCAN}
R.~J. G.~B. Campello, D.~Moulavi, and J.~Sander, ``Density-based clustering based on hierarchical density estimates,'' in \emph{Advances in Knowledge Discovery and Data Mining}, J.~Pei, V.~S. Tseng, L.~Cao, H.~Motoda, and G.~Xu, Eds.\hskip 1em plus 0.5em minus 0.4em\relax Berlin, Heidelberg: Springer Berlin Heidelberg, 2013, pp. 160--172.

\bibitem{Transformer}
A.~Vaswani, N.~Shazeer, N.~Parmar, J.~Uszkoreit, L.~Jones, A.~N. Gomez, L.~Kaiser, and I.~Polosukhin, ``Attention is all you need,'' 2023.

\bibitem{TripletLoss}
E.~Hoffer and N.~Ailon, ``Deep metric learning using triplet network,'' 2018.

\bibitem{FaceNet}
\BIBentryALTinterwordspacing
F.~Schroff, D.~Kalenichenko, and J.~Philbin, ``Facenet: A unified embedding for face recognition and clustering,'' in \emph{2015 IEEE Conference on Computer Vision and Pattern Recognition (CVPR)}.\hskip 1em plus 0.5em minus 0.4em\relax IEEE, Jun. 2015, p. 815–823. [Online]. Available: \url{http://dx.doi.org/10.1109/CVPR.2015.7298682}
\BIBentrySTDinterwordspacing

\bibitem{Adam}
D.~P. Kingma and J.~Ba, ``Adam: A method for stochastic optimization,'' 2017.

\bibitem{GRU}
K.~Cho, B.~van Merrienboer, C.~Gulcehre, D.~Bahdanau, F.~Bougares, H.~Schwenk, and Y.~Bengio, ``Learning phrase representations using rnn encoder-decoder for statistical machine translation,'' 2014.

\bibitem{Bootstrap}
B.~Efron and R.~Tibshirani, ``An introduction to the bootstrap,'' 1994.

\end{thebibliography}


% \end{thebibliography}

\end{document}